\documentclass[aps,twocolumn,superscriptaddress,showpacs]{revtex4}
\usepackage{amsmath}
\usepackage{amssymb}
\usepackage{amsfonts}
\usepackage[dvips]{graphicx}
\usepackage[]{caption}
\usepackage[]{epsfig}
\begin{document}

\title{Depletion forces in non-equilibrium}

\author{J. Dzubiella}
%\email[e-mail address: ] {jd319@cam.ac.uk}
\affiliation{University Chemical Laboratory,
Lensfield Road,
Cambridge CB2 1EW,
United Kingdom}
\author{H. L\"owen}
\affiliation{Institut f{\"u}r Theoretische Physik II,
Heinrich-Heine-Universit\"at D{\"u}sseldorf,
Universit\"atsstra{\ss}e 1, D-40225 D\"usseldorf, Germany}
\author{C. N. Likos}
\affiliation{University Chemical Laboratory,
Lensfield Road,
Cambridge CB2 1EW,
United Kingdom}
\affiliation{Institut f{\"u}r Theoretische Physik II,
Heinrich-Heine-Universit\"at D{\"u}sseldorf,
Universit\"atsstra{\ss}e 1, D-40225 D\"usseldorf, Germany}
\date{\today}

\begin{abstract}
The concept of effective depletion forces between two
fixed big colloidal particles in a bath of small  particles
is generalized to a non-equilibrium situation where
the bath of small Brownian particles is flowing around the big particles 
with a prescribed velocity.
In striking contrast to the equilibrium case,
the non-equilibrium forces violate Newton's third law,
are non-conservative and strongly anisotropic, featuring both
strong attractive and repulsive domains.
\end{abstract}


\maketitle

Colloidal mixtures are excellent model systems to study fundamental 
questions of equilibrium and non-equilibrium phase transitions 
with important physical applications, such as
protein crystallization \cite{Lekk} as well as gel and glass formation 
\cite{Fuchs}. One of the reasons for the pivotal 
role of colloids as model systems lies in the fact that their 
interactions  can be tailored \cite{Pusey}. 
In the theoretical description of colloidal mixtures,
the concept of the effective interaction  has proved 
to be of paramount importance in advancing our
understanding of how to steer the properties of these complex systems
\cite{poon:jpcm:02}. 
It stems from integrating out the coordinates of the small particles,
reducing thereby the number of statistical degrees of freedom
considerably \cite{likos:physrep:01,Springer}. This approach has been
successfully applied to colloid/polymer mixtures 
\cite{AOmodel,Lekkerkerker,jusufi:jpcm:01}, to
binary colloidal mixtures \cite{evans:pre:99,roth:pre:00}
and to charged suspensions \cite{irene:prl:98}. The most widely known
effect is the depletion attraction in strongly non-additive
systems, such as the Asakura-Oosawa model \cite{AOmodel} pertaining to 
colloids and non-adsorbing ideal polymer coils. This attraction
is entropic, i.e., it scales with the thermal energy $k_B T$
and results from a depletion zone of polymer between a pair of nearly touching
colloidal particles, which causes an imbalance of the osmotic pressure
of the polymers acting on the colloidal surface. While the range of
the depletion attraction is typically associated with the radius of gyration 
of the polymers,
its attractive depth is fixed by the polymer concentration,
so that both the range and the depth are in principle tunable.

The crucial limitation of the concept of effective interactions is that it
works, strictly speaking, only in thermodynamic equilibrium 
and can thereafter be applied for the calculations of static quantities,
such as correlation functions and free energies 
(resulting into phase diagrams). In this Letter,
we investigate whether and how the effective interaction can be applied
also in {\it non-equilibrium steady state situations}. In order
to be as transparent as possible, we study a minimal model of
 non-equilibrium for two fixed big colloidal
particles in a quiescent solvent which are exposed to a flowing bath of small
Brownian particles. 
The latter ones are ideal, in full analogy to 
to the equilibrium case of the Asakura-Oosawa model,
and hydrodynamic interactions mediated by the  solvent
are neglected. Despite its simplicity, the model is realized
in colloidal mixtures with weakly interacting depletants
(which justifies the assumed ideality)
which have a small physical volume fraction with 
well-separated excluded volumes
of the big and small particles (which justifies neglecting 
hydrodynamic interactions). Examples include strongly asymmetric
highly charged suspensions with a small physical volume fraction \cite{palberg:jcp:01}
and sterically-stabilized  colloidal particles in a bath of thin rods \cite{koenderink:prl:2003}
or polymer coils \cite{verma:prl:98,rudhardt:prl:98}. Our situation is different to the classical hydrodynamic
problem of two fixed spheres in a Stokes flow of a viscous
incompressible fluid, 
\cite{felderhof:physica:1977,hinsen:jcp:94,jan:book}
in which the force vanishes for zero flow.
More importantly, the physical origin of the depletion 
forces is the high osmotic compressibility
of the bath particles 
complementary to the incompressibility of the viscous fluid mediating
hydrodynamic interactions.

Based on a standard Smoluchowski approach and on
Brownian dynamics computer simulations in non-equilibrium, 
we show that the effective force can be generalized
in a straightforward way to a stationary non-equilibrium state. The 
results for the effective force, however, 
differ  from the
corresponding equilibrium case both qualitatively and quantitatively.
First of all, the non-equilibrium force field (as a function of the mutual
distance between the particles) is non-conservative and violates Newton's
third law, implying that the concept of an effective interaction potential
does not hold in non-equilibrium. 
Second, when the small particles are brought to flow, the 
equilibrium depletion attraction
gets strongly anisotropic
favoring an alignment of the colloidal pair along the flow
direction. Conversely,
there is mutual effective
repulsion between the colloidal pair perpendicular to the drift
direction of the bath due to a compression of the small particles between the
colloids. Furthermore, the range of the depletion interaction in the
drifting case is larger than that in equilibrium. 

Equilibrium depletion forces
can be measured in real colloidal samples with high 
accuracy \cite{ koenderink:prl:2003,verma:prl:98,rudhardt:prl:98,
bechinger:prl:99, crocker:prl:99}.
Our predictions in non-equilibrium can be tested in experiments 
of binary colloidal mixtures as well.
A drifting bath is generated 
either by gravity \cite{philipse:97, aarts:jpcm:03} 
or by an electric or magnetic field \cite{palberg:jcp:01}
and the big particles can be  fixed by optical tweezers \cite{crocker:prl:99, yodh:01}. 
The measurement of the relative displacement with respect to the tweezer's
position of the big particles
provides direct access to the effective non-equilibrium forces.
  
\begin{figure}
  \begin{center}
\includegraphics[width=8.5cm,angle=0.,clip]{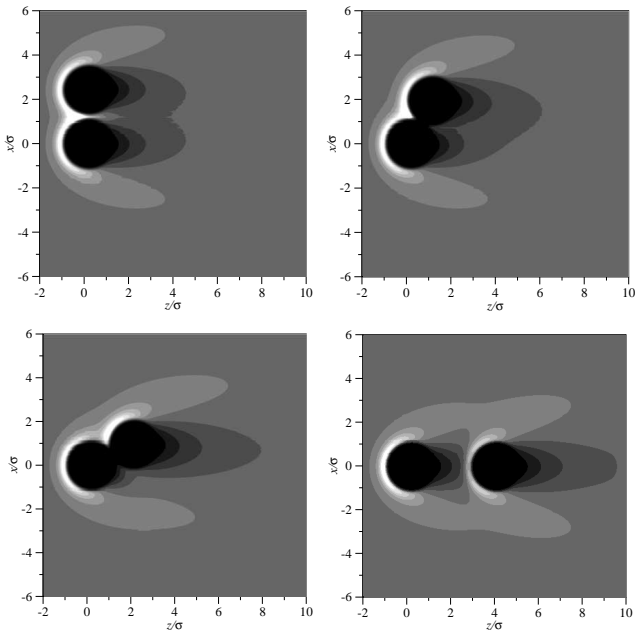}
    \caption{Steady-state contour density fields of non-interacting
      Brownian particles, driven around two stationary colloidal spheres,
      on the $x-z$ plane where the centers of the spheres are located. The
      distributions are obtained from the superposition approximation, 
      Eq.\ (\ref{dsa:eq}).
      Bright regions show high densities while
      low density regions are plotted dark. One colloid is placed
      at the origin, ${\bf R}_1 = 0$, and the position ${\bf R}_2$
      of the other one is:
      (a) ${\bf R}_2/\sigma=(2,0,0)$; (b) ${\bf R}_2/\sigma=(1.5,0,1)$;
      (c) ${\bf R}_2/\sigma=(1,0,2)$; (b) ${\bf R}_2/\sigma=(0,0,4)$.
      In this figure, the bath particles drifting from the left to the
      right with a speed $v^{*} = 5$.}
\label{profiles:fig}
\end{center}
\end{figure}
In our theoretical model we consider two fixed big particles
in a bath of small Brownian particles
that drift with velocity ${\bf v} = v{\bf{\hat z}}$.
The bath has a bulk number density $\rho_0$.
While there is no interaction between the
bath particles, the
interaction between the big and small particles 
is described with an exponential,
spherically symmetric potential $V({\bf r})$ given by
\begin{eqnarray}
\nonumber
V({\bf r})=V(r)=V_0 \exp\left[-(r/\sigma)^{n}\right],
\end{eqnarray}
\begin{figure}
\vspace{-0.0cm}
\begin{center}
    \includegraphics[width=8.5cm, angle=0., clip]{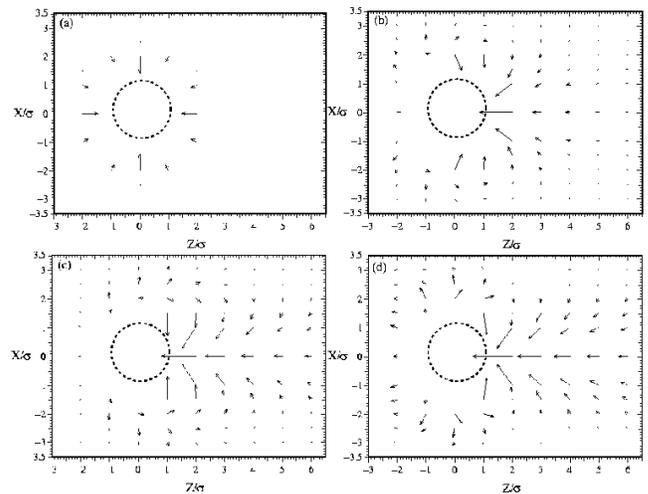}
    \caption{The depletion force ${\bf F}_{{\rm depl},2}$ (arrows)
    acting on colloid 2, when colloid 1
    is placed at the origin (dashed circle), as the Brownian
    bath particles drift from the left to the right with speed $v^{*}$.
    Forces are plotted in the $X-Z$
    plane where the centers of the colloids are located. Results are
    shown for drift velocities (a) $v^{*}=0$ (equilibrium), (b) $v^{*}=1$, and
    (c) $v^{*}=5$.
    The results in (a)-(c) are calculated theoretically from
    Eqs.\ (\ref{dsa:eq}) and (\ref{total:eq}).
    Plot (d) shows the force for $v^{*}=5$
    resulting from Brownian Dynamics computer simulations. Different
    scales for the length of the vectors are used for different values
    of $v^{*}$. The magnitude of the force vector located at
    $(X,Z)=(0,2)$ is $\beta F\sigma=1.22$ in (a), 2.38 in (b), 8.04 in
    (c), and 9.16 in (d). Furthermore, $\rho_0 =1/\sigma^{3}$.}
  \label{fields:fig}
\end{center}
\end{figure} 
where ${\bf r}$ is the vector separating the bath particle with
the center of the colloid, $r$ is the magnitude of ${\bf r}$ and
$V_0=10 k_{B}T$. 
In this work we focus on an exponent $n=6$, on one hand to deal with a
well-defined `hard' interaction diameter of the colloidal sphere, $\sigma$, on the other
hand to circumvent numerical inconveniences due to huge forces at
$r=\sigma$ for high exponents $n$. 

In what follows, we first discuss a theoretical Smoluchowski
approach for a single driven colloid. Adopting
a dynamical superposition approximation (DSA), we
then address the case of two colloids and afterwards we compare
the theoretical predictions arising from the DSA with Brownian dynamics (BD) 
computer simulations, finding
 good agreement between the two approaches. In detail,
for the case of a {\it single} big colloidal particle, the Smoluchowski 
equation \cite{jan:book} for the steady state  density
field $\rho({\bf r})$ 
of the bath particles reads 
as $\vec\nabla\cdot{\bf j}({\bf r}) = 0$ with the current density field
${\bf j}({\bf r})$ composed
of 
\begin{eqnarray}
\nonumber
\label{eq:4} 
{\bf j}({\bf r})={\frac{\Gamma}{\beta}}\vec\nabla\rho({\bf r}) +
                 \Gamma\rho({\bf r}) \vec\nabla V({\bf r})
                 +{\bf v}\rho({\bf r}),
\end{eqnarray}
where 
$\beta=1/k_{B}T$ is the inverse thermal energy and $\Gamma$ 
the mobility of the bath particles in the solvent.
As the problem possesses azimuthal symmetry,
these equations can be readily numerically solved in cylindrical
coordinates.
Let us define the resulting density profile of the bath
particles around one big
colloid as $\rho^{(1)}({\bf r}-{\bf R};{\bf v})$
with ${\bf R}$ denoting the position of the colloid.
Spatial homogeneity dictates that
$\rho^{(1)}$  depends only on the difference
${\bf r}-{\bf R}$. The force ${\bf F}_{\rm d}$ acting on the big
particle due to the drifting bath is then obtained as ${\bf F}_{\rm
  d}=-\int {\rm d}^{3}r \rho^{(1)}({\bf r}){{\bf \nabla}_{}}V(|\bf r|)$.

In the presence of {\it two} colloids, the corresponding quantity
of interest is the bath density profile
$\rho^{(2)}({\bf r},{\bf R}_1,{\bf R}_2;{\bf v})$
that depends on the locations ${\bf R}_i$ of the colloids, ($i = 1, 2$),
and parametrically on the velocity ${\bf v}$. 
In this case, the numerical solution
of the Smoluchowski equation is more difficult such that a direct
 particle-resolved computer simulations is more appropriate.
However, let us motivate a simple superposition approximation.
As the bath particles are
noninteracting, in the case of thermodynamic equilibrium,
$v = 0$, $\rho^{(2)}({\bf r},{\bf R}_1,{\bf R}_2;0)$ can
be exactly factorized as the product of the density
fields $\rho^{(1)}({\bf r}-{\bf R}_i;0)$ divided by $\rho_0$.
In what follows, we employ the same factorization ansatz also for
the non-equilibrium situation, $v \ne 0$. Thereby, we introduce the
dynamical analog of the superposition approximation
that was first introduced by Attard \cite{attard:jcp:89} for
{\it equilibrium} problems dealing with {\it interacting} particles,
in order to calculate static depletion forces.
Hence, in this
{\it dynamical superposition approximation} (DSA), we write:
\begin{eqnarray}
\rho^{(2)}({\bf r},{\bf R}_1,{\bf R}_2;{\bf v})
\approx \rho^{(1)}({\bf r}-{\bf R}_1;{\bf v})
              \rho^{(1)}   ({\bf r}-{\bf R}_2;{\bf v})/\rho_0.
\label{dsa:eq}
\end{eqnarray}

Results from the DSA are shown in Fig.\ \ref{profiles:fig}, where we plot
the density distribution in the $x$-$z$ plane, on which the centers of
both colloids are located. The profiles are plotted for four different
relative positions of the two colloids and for a dimensionless drift velocity
$v^{*}=5$, defined 
as $v^{*}=\beta\sigma v/\Gamma$. The density profiles are strongly
anisotropic and show long-ranged low-density regions in the wake
region behind the spheres. The extension 
of the concept of the effective interaction to a steady-state non-equilibrium
situation can now be performed via the definition of the effective {\it force}
which is a statistical average over the forces of the small bath
particles exerted onto the big ones \cite{likos:physrep:01}. 
Generalizing to a steady-state average in non-equilibrium,
the effective force reads as \cite{footnote}:
\begin{eqnarray}
{\bf F}_i = -\int\,{\rm d}^{3}r\,\rho^{(2)}({\bf r},{\bf R}_1,{\bf R}_2;{\bf v})
             \vec\nabla_{{\bf R}_i} V(|{\bf R}_i-{\bf r}|). 
\label{total:eq}
\end{eqnarray}

\begin{figure}
\vspace{-0cm}
\begin{center}
\includegraphics[width=8.5cm,angle=0.,clip]{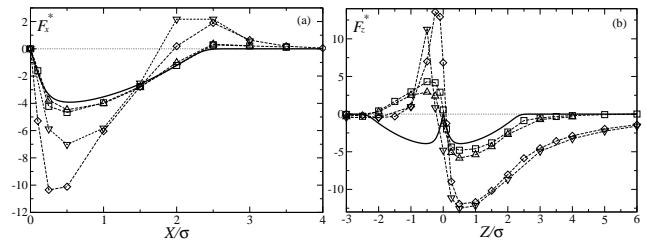}
    \caption{Depletion forces as in Fig.\ \ref{fields:fig}, but now we plot the
     projection of the the depletion force 
     $F_{x}^{*}(X)=\beta\sigma{\bf F}_{{\rm depl},2}\cdot{\bf\hat X} $
     for fixed $Z=0$ in the $X$ direction (a)
     and the projection of the depletion force
     $F_{z}^{*}(Z)=\beta\sigma{\bf F}_{{\rm depl},2}\cdot{\bf\hat z} $
     for fixed $X=0$ in $Z$-direction. 
     The solid line is the exact result for $v^{*}=0$ and the symbols are
     results from the DSA for $v^{*}=1$ (squares) and $v^{*}=5$
     (diamonds), as well as simulation results for $v^{*}=1$
     (triangles up) and $v^{*}=5$ (triangles down).}
  \label{fxfz:fig}
\end{center}
\end{figure}
The force has been calculated in theory by employing the DSA,
Eq.\ (\ref{dsa:eq}), for various different configurations of
the colloids and drift velocities, including the equilibrium
case, $v=0$, in which the results are exact. In this way, 
a comparison between the static and dynamic depletion force
${\bf F}_{{\rm depl},i}$ acting on the
$i$-th particle can be obtained. The latter is defined as 
the force ${\bf
  F}_{{\rm depl},i} = {\bf F}_i-{\bf F}_{\rm d}$, 
relative to the constant the drift force ${\bf F}_{\rm d}$
acting on a single big particle 
($\beta  F_{\rm d}\sigma=2.96$ for $v^{*}=1$ and
$\beta  F_{\rm d}\sigma=11.76$ for $v^{*}=5$).
We accompanied the DSA with standard Brownian
Dynamic simulations \cite{Allen}, in which the Langevin equations for each
bath particle are solved iteratively. Here, the bath particles are
driven by an external force ${\bf v}/\Gamma$ for a prescribed flow
velocity ${\bf v}$.
The box lengths
in all simulations 
were $L_{x}=L_{y}=10\sigma$ and $L_{z}=20\sigma$ and 
$N=2000$ particles were simulated, applying periodic
boundary conditions in all directions.

Results for the averaged force field acting on a colloid placed 
at the position ${\bf R}_2$ away from another colloid at the
origin (${\bf R}_1 = 0$), are depicted in Fig.\ \ref{fields:fig}.
A comparison between Figs.\ \ref{fields:fig}(c) and \ref{fields:fig}(d),
(DSA vs.\ BD-simulation), shows that the former captures the
features of the non-equilibrium depletion force quantitatively,
even for drifting speeds as large as $v^{*} = 5$. 
Whereas for the case of equilibrium, Fig.\ \ref{fields:fig}(a),
we obtain spherically symmetric {\it attractive} 
depletion forces, the dynamical
depletion force shows new qualitative features:
First, 
as is clear from inspection 
of Fig.\ \ref{fields:fig}, the dynamical depletion forces violate 
Newton's third law. 
Second, the violation of the ``${\rm action}={\rm reaction}$''-principle
implies a non-conservative force. Indeed, if a putative depletion potential
$V_{\rm depl}$ existed in the dynamical case at hand, translational
invariance of space (which is not broken by the steady flow) would
imply $V_{\rm depl} = V_{\rm depl}({\bf R}_1 - {\bf R}_2)$. As
${\bf F}_{{\rm depl},i} =
-\vec\nabla_{{\bf R}_i} V_{\rm depl}({\bf R}_1 - {\bf R}_2)$, $i=1,2$,
this would have the consequence
${\bf F}_{{\rm depl},1} = -{\bf F}_{{\rm depl},2}$. Third,
the depletion force shows strong anisotropy,
becoming attractive as the colloid starts positioning 
itself `behind' the one at the origin and
its magnitude (in the attractive case) and its range 
are vastly larger for $v \ne 0$ than in equilibrium.

In order to better quantify the features of the depletion force 
in non-equilibrium, we
show in Fig.\ 3 plots of the projected depletion force at fixed
$X = 0$ (particles fully aligned along the flow direction)
and at fixed $Z = 0$ (particles fully aligned perpendicular to the
flow direction). The depletion force exhibits a  
{\it repulsive barrier} in the $X$-direction due to compression of the
small particles between the colloids, in qualitative difference to the 
{\it purely attractive}
equilibrium case $v=0$. In the $Z-$direction, on the other hand, 
there is much stronger
attraction for $Z>0$ due to a long low density region in the wake of
the colloids but a marked  repulsive barrier for $Z<0$. Furthermore,
the range of the depletion force is significantly 
larger than in equilibrium. 
Notice that we show forces also for the case of completely
overlapping colloids \cite{footnote}, 
$-2.0\lesssim X/\sigma\lesssim 2.0$ and $-2.0\lesssim Z/\sigma\lesssim 2.0$.
For ${\bf R}_1 = {\bf R}_2$, the depletion force
on both colloids is obviously the same, 
${\bf F}_{{\rm depl},1} = {\bf F}_{{\rm depl},2}$ 
but, in contrast to the equilibrium case, it does not vanish. 
We can finally deduce from Fig.\ \ref{fxfz:fig} 
that the computer simulation results
are in good quantitative agreement with the theory up
to drifting  speeds of  $v^{*}=1$, but 
for large drifting speeds as high as $v^*=5$ deviations
near the maxima and minima are getting more pronounced.

In conclusion, we have generalized the depletion concept to
a simple non-equilibrium situation of two big particles
in a Brownian flow of small ideal particles.
The effective forces averaged in the steady state are
strikingly different from those in equilibrium:
they are strongly anisotropic with both attractive
and repulsive domains and 
violate Newton's third law, in strong resemblance to the
so-called ``social forces'' that have been applied
extensively in the modelling of pedestrian and automotive
traffic flows \cite{helbing:book}.

Although the present study has focused on noninteracting
bath particles, the extension to interacting ones can be
carried out within the framework of the 
recently proposed dynamical density
functional theory (DDFT) \cite{marconi:jcp:99}.
We expect that 
interaction effects will weaken the strength of the dynamical
depletion forces, as they will tend to smoothen out regions
of spatial inhomogeneity of the bath density that give rise
to these forces in the first place. Finally, we remark that
the strong attractive forces tending to align the colloids
in the direction of the incoming flow can be regarded as
precursors of the laning transition 
that has been analyzed
recently for concentrated colloidal 
mixtures \cite{joe:laning1}. 

We thank R. Finken, G. N\"agele, and H. Stark for useful discussions.
Financial support from the EPSRC and from the DFG through the SFB-TR6
are gratefully acknowledged. C.N.L.\ has been supported by the DFG
through a Heisenberg Fellowship
and wishes to thank the Department of
Chemistry, University of Cambridge, and Prof.\ J.-P. Hansen for their
hospitality.


\begin{thebibliography}{99}
\bibitem{footnote}
We do not need to specify the direct
interaction between the two big colloids,
$V_{\rm cc}(|{\bf R}_1-{\bf R}_2|)$, since we focus exclusively on the
{\it additional}, dynamical depletion forces
which are mediated by the flow of the bath particles.
\end{thebibliography}
\end{document}